# FRAME SIZE OPTIMIZATION USING A MACHINE LEARNING APPROACH IN WLAN DOWNLINK MU-MIMO CHANNEL

Lemlem Kassa [1], Jianhua Deng [2], Mark Davis[3] and Jingye Cai [4]

[1,2,4]School of Information and Software Engineering, University of Electronic Science and Technology China (UESTC), Chengdu 610054, China
lemlem.kassa@aastu.edu.et; jianhua.deng@uestc.edu.cn; jycai@uestc.edu.cn.
[3] Communication Network Research Institute (CNRI), Technological University, D08 NF82 Dublin, Ireland
mark.davis@tudublin.ie

## ABSTRACT

*The IEEE 802.11ac/n introduced frame aggregation technology to accommodate the growing traffic demand and increase the performance of transmission efficiency and channel utilization. This is achieved by allowing many packets to be aggregated per transmission which realized a significant enhancement for the throughput performance of WLAN. However, it is difficult to efficiently utilize the benefits of frame aggregation in the downlink MU-MIMO channels as stations have heterogeneous transmission demands and data transmission rates. As a result of this, wasted space channel time will occur which degrades transmission efficiency. In addressing these challenges, the existing studies have proposed different approaches. However, most of these approaches did not consider a machine-Learning based optimization solution. The main contribution of this paper is to propose a machine-learning-based frame size optimization solution to maximize the system throughput of WLAN in the downlink MU-MIMO channel. In this approach, the Access Point (AP) performs the maximum system throughput measurement and collected frame size-system throughput patterns which contain knowledge about the effects of traffic patterns, channel conditions, and number of stations (STAs). Based on these patterns, our approach uses a neural network to correctly model the system throughput as a function of the system frame size. After training the neural network, we obtain the gradient information to adjust the frame size. the performance of the proposed Machine learning (ML) approach is evaluated over the FIFO aggregation algorithm under the effects of heterogenous traffic patterns for VoIP and video applications, channel conditions, and number of stations.*

## 1. INTRODUCTION

Due to the advancement of wireless technologies, IEEE 802.11 based networks are becoming more popular, and different technologies have been introduced to improve throughput performance. Multi-user multiple-input multiple-output (MU-MIMO) is among the ones at the physical layer introduced by IEEE 802.11ac standard to accommodate the increasing demand of high data transmission rate by allowing a single Access Point (AP) supports simultaneous transmission up to a maximum of eight users at a time [1,2]. This is one of the most crucial technologies that has driven wireless local area networks (WLANs) toward the gigabit era. Moreover, the wireless medium has a high overhead in terms of bytes that can be higher than the actual payload. To amortize these overheads such as the Medium Access control (MAC) and physical (PHY) headers, acknowledgments (ACK), backoff time, and inter-frame spacing, the standard also introduced a frame aggregation scheme which has contributed to a high data throughput by combining multiple frames, also known as MAC Service Data Units (MSDUs), into a single transmission unit [1]. The performance of WLAN depends on different performance factors such as frequency channel, modulation, and coding schemes, transmitter power, etc. at the PHY layer, and retry limit, frame size, contention window size, maximum number of backoffs, etc. at the MAC layer have a significant impact on the performance of WLAN. Optimizing these parameters would improve the system performance of WLAN. Frame size optimization is the main concern of this study. If a wireless frame size is large, a bit error would destroy the whole frame thus the frame success rate decreases, and also the throughput performance degrades [3]. On the other hand, if the frame size is shorter, the overhead frames such as MAC and PHY headers occupy a large portion of the transmitted frame thus degrade the transmission efficiency [3]. IEEE 802.11 standard specified a constant length aggregation strategy regardless of the traffic pattern and channel conditions. This contributes to the reduction of channel access overhead. However, utilizing the maximum aggregation size may not be optimal in all channel conditions and traffic patterns because it may lead to an increase in the delivery of error frames and retransmissions [4]. This phenomenon particularly degrades the performance of WLAN in downlink MU-MIMO channel when streams have heterogeneous traffic demands such that variable transmission rate among spatial streams causes space channel time. Therefore, determining the optimal frame size is significant to improve the system throughput of WLAN [3,4].

The development of smart devices, mobile applications, and wireless users' interaction with wireless communication systems are significantly increased and caused a massive amount of traffic being generated in the communication network [5]. These challenges have pushed the wireless networking industry to seek innovative solutions to ensure the required network performance. On the other hand, the release of the new IEEE 802.11 standards such as IEEE 802.11ax and IEEE 802.11ay, and 5G technologies expand the set of available communication technologies which compete for the limited radio spectrum resources in pushing for the need of enhancing their coexistence and more effective use of the scarce spectrum resources. In response to these performance demands, more recently, Machine Learning (ML) approaches have started to attract significant attention and are increasingly used in the context of wireless communication systems to generate a self-driven networks that can configure and optimize themselves by reducing human interventions [6-10]. The developments of mobile edge caching and computing technologies also made it possible for base stations (BSs) to store and analyze human behavior for wireless communication [8-10]. Therefore, the evolution toward learning-based data-driven network systems helps to develop and realize many of the promising benefits obtained from ML. ML is used to develop advanced approaches that can autonomously extract patterns and predict trends based on environmental measurements and performance indicators as input. Such patterns can be used to optimize the parameter settings at different protocol layers, e.g., PHY, and MAC layers [5, 6]. Several frame size optimization schemes are proposed to improve the throughput performance of WLAN. For instance, [11] proposed the adaptive frame size estimation scheme depending on the channel condition to improve the throughput performance of WLAN in the error-prone channel based on the Extended Kalman Filter. By studying the relationship between throughput and frame size, [12] illustrated that throughput is a monotonically increasing function of the frame size, i.e., the larger the frame size, the better the throughput. However, these approaches do not provide a machine-learning based optimization solution and the algorithms are not applicable in IEEE 802.11 MU-MIMO enabled WLAN. In considering the channel condition and contention effects in WLAN, [13] proposed a machine learning-based adaptive approach for frame size optimization, however, this approach is not applicable in MU-MIMO enabled WLAN. The main contribution of this paper is to propose a machine-learning based adaptive algorithm to optimize the frame size that would maximize the system throughput in the WLAN downlink MU-MIMO channel considering the effect of channel condition heterogeneous traffic patterns and number of stations. Thus, our proposed ML approach is significant as it can autonomously extract patterns and predict trends based on environmental measurements and performance indicators as input.

The rest of the paper is organized as follows, in Section 2, we introduce related works about the frame aggregation schemes and the performance challenges of multi-user transmissions in the WLAN downlink MU-MIMO channel. A detailed problem description of the proposed machine-learning approach is given in Section 3. In Section 4, results and discussions are presented to evaluate the performance of the proposed approach under various channel conditions, traffic models, and number of stations. Finally, the conclusions are given in Section 5.

## 2. RELATED WORK AND OUR MOTIVATION

In this section, some previous works and the effects of frame size determination approaches on the performance of WLAN are discussed mainly focusing on the downlink MU-MIMO channel.

### 2.1. Related Work

Frame size optimization problem has been studied by several researchers for IEEE 802.11 networks. For instance, employing a specific procedure of dynamically adjusting the frame size, [11] proposed a method that deals with frame size estimation based on the extended Kalman Filter for saturated networks. They derive the mathematical equation of throughput, which is a function of the frame size. The optimal frame size is obtained using differential calculus. Bianchi's Markov chain model studied the relationship between the throughput and frame size, in IEEE 802.11 WLANs [12]. However, the assumption of this work is ideal channel which is unrealistic. According to the simulation results, the throughput increases with the frame size, i.e., the larger the frame size, the better the throughput. A machine learning-based frame size optimization approach in considering both channel conditions and contention effects of users is proposed by [13]. According to the simulation results, the frame size optimization is effectively achieved to maximize the throughput performance of WLAN. However, this approach does not support the frame aggregation mechanism and the algorithm is not suitable for IEEE 802.11 MU-MIMO enabled WLAN. An adaptive algorithm for frame size optimization is proposed by [14] which allows an ARQ protocol to dynamically optimize the packet size based on estimates of the channel bit-error. The main strategy of this study is, to make estimates of the channel bit-error-rate, they consider the acknowledgment history, thus based on that the optimal packet size can be determined. However, this approach is not suitable for IEEE 802.11 WLAN environments.

Moreover, some studies contributed frame size aggregation schemes in WLAN downlink MU-MIMO channels to enhance the throughput performance [15-22]. The algorithm in [15] proposed a new approach aiming to enhance the system throughput performance of WLAN employing a dynamic adaptive aggregation selection scheme to determine the optimal length of the frame size in downlink MU-MIMO transmission. The effects of heterogeneous traffic demand among spatial streams are considered under the assumption of ideal channel. According to the simulation results, the maximum performance of system throughput performance and channel utilization is achieved. By extending the work of [15], an adaptive frame aggregation algorithm is proposed by [16] in considering the effect of transmission error. Moreover, a data frame construction scheme called DFSC [17] proposed to find the length of a Multi-User (MU) frame aiming to maximize the transmission efficiency by considering the status of buffers and transmission bit rates of stations in both uplink and downlink multiuser transmissions. However, this work did not consider the effect of channel error that could reduce the transmission performance due to excessive retransmissions of frames received in error. A frame size-based aggregation scheme is proposed by [18] where the authors demonstrated that both the queueing length and number of active nodes have significant impacts on the system throughput performance. The main approach of this paper is to generate the same frame length in all spatial streams that could maximize the system throughput performance. [19] proposed a novel method to determine frame aggregation size in MU-MIMO channel to improve channel utilization in considering delay data frames wait in transmission queues. Some works in the literature have also been studied focusing on the padding problem. According to [20,21], they improved transmission efficiency in the downlink MU-MIMO channel by replacing padding bits with data frames from other users in one stream to fill the space of frame padding violating the rules of MU transmissions. However, these approaches increased the complexity of both the transmission and reception process in wireless communication which requires modification of the standard to allow the transmission to multiple destinations within a special stream. A frame duration-based frame aggregation scheme is proposed in [22] by employing user selection criteria by providing high priority to the MT expecting high throughput in the next MU-MIMO transmission and having a large amount of data while reducing signaling overhead. The main approach of this study is, by equalizing the transmission time of all spatial streams in all MTs according to their Modulation and Coding (MCS) level they could achieve the maximum system throughput and minimize space channel time in WLAN the downlink MU-MIMO channel. Although all the above proposals contributed several schemes to enhance the performance of WLAN, none of them has proposed a machine-learning based optimization solutions. To the best of our knowledge, there is little research explored with the use of ML techniques to tackle frame size optimization problems in WLAN. In contrast to these approaches our work attempt to propose a machine learning-based adaptive approach for frame size optimization in the WLAN downlink MU-MIMO channel.

## 2.2. Motivation for this Work

The dynamic adaptive frame aggregation selection scheme can maximize the system throughput performance of WLAN while enhancing system throughput performance by minimizing space channel time. However, this approach does not consider a machine learning-based optimization solution. The motivation of this work comes with the aim of extending the previous work [16] to contribute a machine learning base adaptive approach for frame size optimization to maximize the system throughput performance of WLAN in the downlink MU-MIMO channel. Thus, we can generate a self-driven networks that can configure and optimize themselves by reducing human interventions. Moreover, for the growing diversification of services, users, and the constantly changing channel and traffic dynamics in a networking system, a ML solution is relevant and should be adopted in more effective ways to speed up the decision-making process [4–7].

## 3. PROPOSED APPROACH

In this topic, the problem definitions and the proposed machine learning approach are discussed.

### 3.1. Problem Definition

In this paper, we tackle the frame-size optimization problem using a machine-learning-based adaptive approach in considering the effects of traffic patterns, channel conditions, and number of stations in WLAN downlink MU-MIMO. In this approach, the simulation environment proposed by [16] is used to collect the "frame size–system throughput "patterns. The frame size represents the average offered traffic load in [Mbps] generated in the system by employing different traffic models (Pareto, Weibull, or fractional Brownian Motion (fBM)) [15,16]. System throughput is defined as the average system data rate at which the AP can successfully transmit to all receiving stations. The collected patterns contain knowledge about the effects of traffic patterns, channel conditions, and number of stations. Suppose *frm* is the frame size and *Thr* is the corresponding throughput. Based on these patterns, our approach uses neural networks to build the knowledge and accurately model the throughput *Thr* as a function with respect to the frame size *frm*. The neural network is a good approach to model a system effectively that may contain some noise [13,23]. Thus, after the knowledge building, we obtain the gradient information from the neural networks and adaptively adjust the frame size based on the gradient information. In the formation of frame-size optimization problem, the throughput *Thr* is a complex function of the frame size *frm* under some channel conditions and traffic patterns and number of stations, i.e., *Thr= f(frm)*. The function *f* varies with the channel conditions, traffic pattern and number of stations in the network. Therefore, how the throughput *Thr* can be maximized by optimizing the frame size *frm* is the main focus of the problem in this study.

$$frm_{Opt} = \underset{frm}{\mathrm{argmax}}\, Thr = \underset{frm}{\mathrm{argmax}}\, f(frm) \qquad (1)$$

Therefore, the goal of this approach is to choose the optimal frame size that would maximize the objective function *Thr*. The objective function of this optimization problem defined as the throughput *Thr= f(frm)*. However, due to the dynamic effects of channel conditions and traffic patterns, it is difficult to analyze and obtain an accurate throughput function *f(frm)* in all network conditions. Thus, we solved such an optimization problem by adopting the well-known gradient ascent algorithm [24]. Such that the local maximum of the throughput function *Thr = f(frm)* can be found by adaptively adjusting frame size *frm* using gradient ascent, by taking steps that are proportional to the gradient. Suppose that, at the $n^{th}$ time of adjustment, the frame size is *frm(n)*, and the throughput is *Thr(n)*. At the next time of adjustment, the frame size *frm* is set as:

$$frm(n+1) = frm(n) + \Delta frm(n) \qquad (2)$$

Where $\Delta frm(n)$ depends on the gradient of the estimated throughput $Thr(n)$ with respect to $frm(n)$, i.e.,

$$\Delta frm(n) = \mu \frac{\partial Thr(n)}{\partial frm(n)} \qquad (3)$$

The parameter μ is a variable adjustment rate heuristically selected for different network scenarios. Then, to solve the gradient problem *(∂Thr(n))/(∂frm(n))*, a machine-learning-based adaptive approach is elaborated in the following sub-section.

## 3.2. The Proposed Machine- learning-based Adaptive Solution

Machine Learning (ML) is an innovative solution that can autonomously extract patterns and predict trends based on environmental measurements and performance indicators as input to provide self-driven intelligent network systems that can configure and optimize themselves. Under the effects of heterogeneous traffic demand among users and varying channel conditions in WLAN downlink MU-MIMO channels, achieving the maximum system throughput performance is challenging. Online learning (also called incremental learning) and offline learning (or batch learning) are types of learning strategies in machine learning [26]. In online learning, the algorithm updates its parameters after learning from each individual training instance i.e., it is fed with individual data or mini-baches [25,26]. This allows the learning algorithm keep learning on the fly, after being deployed as new data arrives. The weight changes in online learning made at a given stage depend specifically only on the current training instance being presented and possibly on the current state of the model. When an online model has learned from new data instances, it no longer needs to use them and can therefore discard them. This can save a huge amount of memory space. Whereas traditional machine learning is performed offline using offline learning which is the opposite of online learning [26]. On the contrary, in offline learning, the learning algorithm updates its parameters after consuming the whole batch, and the weight changes depending on the whole (training) dataset, defining a global cost function [26]. Therefore, in this study to cope with the effects of time-varying channel conditions and heterogeneous traffic patterns, the online machine learning strategy is employed to achieve the data collection, knowledge building, and frame-size adjustment kept online.

The proposed Multi-Layer Perception (MLP) ML approach consists of one hidden layer with four neurons and an output layer. The backpropagation algorithm only consists of two passes: 1) a forward pass and 2) a backward pass [27]. To obtain the gradient information $\partial Thr(n)/\partial frm(n)$ which is used to adjust the frame size, we add a third pass, i.e., the tuning pass as shown in Figure 1. In the proposed MLP approach, the backpropagation algorithm is used to adjust the network and minimize the error between the actual response and the desired (target). The detailed description of the tuning pass is provided as follows including a summary of notations in Table I.

### 3.2.1. Tuning Pass Strategies

The diagram shown in Figure 1 illustrated the signal flow of the tuning pass in the machine learning model to estimate the gradient $\frac{\partial \widetilde{Thr}(n)}{\partial frm(n)}$, and the key to adjusting the frame size to maximize the throughput. The initial weight is denoted as $w_{ij}^i$ in the neural network is randomly chosen. The synaptic weights that have been well adjusted in the backward pass are set as fixed in the tuning pass. An adaptive learning rate is adopted to improve the convergence speed [25].

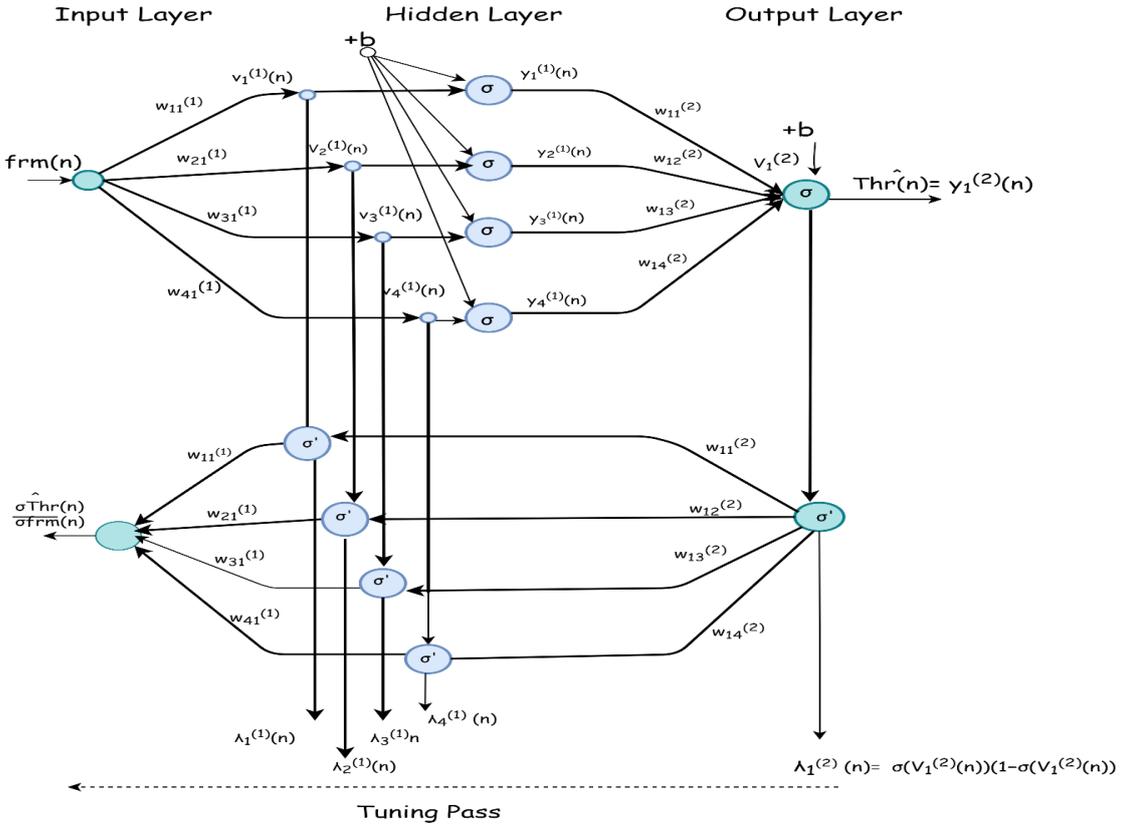

Figure 1. Flow chart of the proposed machine learning approach consisting of tuning pass, which depicts the derivation of the local gradients and the gradient for frame size adjustment.

In the following discussion, the procedure how the estimated gradient $\frac{\partial \widetilde{Thr}(n)}{\partial frm(n)}$ can be obtained is presented. Considering the hidden layer, the local gradient $\lambda_j^{(l)}(n)$ for the tuning pass is defined as follows:

$$\lambda_j^{(l)}(n) = \frac{\partial \widetilde{Thr}}{\partial v_j^{(l)}(n)} \tag{4}$$

Where $v_j^{(l)}$ in equation (4) is the weight sum of synaptic input plus bias of neuron $j$ in layer $l$. Similarly, considering the output layer, the local gradient $\lambda_1^{(2)}(n)$ is defined as follows:

$$\lambda_1^{(2)}(n) = \frac{\partial \widetilde{Thr}(n)}{\partial v_1^{(2)}(n)} = \partial'(v_1^{(2)}(n)) = \partial(v_1^{(2)}(n))(1 - \partial(v_1^{(2)}(n))) \tag{5}$$

While considering the hidden layer, the local gradient $\lambda_j^{(l)}(n)$ can be expressed as follows using the chain rule:

$$\lambda_j^{(1)}(n) = \frac{\partial \widetilde{Thr}(n)}{\partial v_j^{(1)}(n)} = \frac{\partial \widetilde{Thr}(n)}{\partial v_1^{(2)}(n)} \cdot \frac{\partial v_1^{(2)}(n)}{\partial y_j^{(1)}(n)} \cdot \frac{\partial y_j^{(1)}(n)}{\partial v_j^{(l)}(n)}$$

$$\lambda_1^{(2)}(n) \cdot w_{1j}^{(2)}(n) \cdot \partial'(v_j^{(1)}(n)) \tag{6}$$

Therefore, using the results (5) and (6), the gradient can be written as follows:

$$\frac{\partial \widetilde{Thr}(n)}{\partial frm(n)} = \frac{\partial \widetilde{Thr}(n)}{\partial v_1^{(2)}(n)} \cdot \frac{\partial v_1^{(2)}(n)}{\partial frm(n)}$$

$$= \lambda_1^{(2)}(n) \cdot \frac{\partial v_1^{(2)}(n)}{\partial frm(n)} \tag{7}$$

Where $v_1^{(2)}(n)$, can be defined as $v_1^{(2)}(n) = \sum_{i=1}^{4} w_{1j}^{(2)}(n) \cdot y_j^{(1)}(n)$. Thus, the second term at the rightmost side of equation (7) can be written as:

$$\frac{\partial v_1^{(2)}(n)}{\partial frm(n)} = \sum_{i=1}^{4} w_{1j}^{(2)}(n) \cdot \frac{\partial y_j^{(1)}(n)}{\partial frm(n)}$$

$$= \sum_{i=1}^{4} w_{1j}^{(2)}(n) \cdot \frac{\partial y_j^{(1)}(n)}{\partial v_j^{(1)}(n)} \cdot \frac{\partial v_j^{(1)}(n)}{\partial frm(n)}$$

$$= \sum_{i=1}^{4} w_{1j}^{(2)}(n) \cdot \partial'(v_j^{(1)}(n)) \cdot w_{j1}^{(1)}(n)$$

$$= \sum_{i=1}^{4} \lambda_1^{(2)}(n) \cdot w_{1j}^{(2)}(n) \cdot \partial'(v_j^{(1)}(n)) \cdot w_{j1}^{(1)}(n)$$

$$= \sum_{i=1}^{4} \lambda_j^{(1)}(n) \cdot w_{j1}^{(1)}(n) \tag{8}$$

Therefore, the gradient $\frac{\partial \widetilde{Thr}(n)}{\partial frm(n)} = \sum_{i=1}^{4} \lambda_j^{(1)}(n) \cdot w_{j1}^{(1)}(n) \tag{9}$

The derivation of the local gradients at each layer and the gradient $\frac{\partial \widetilde{Thr}(n)}{\partial frm(n)}$ is depicted in Figure 1.

Based on the result from equation (9), the frame size *frm* is adjusted as shown in the equations (2) and (3).

In general, Figure 2 illustrates the basic components and flow of the proposed ML approach. As shown in the figure, once the AP collects the instant learning dataset from the simulation experiment [16] as a pattern of frame size-system throughput, the neural network performs the training and adjusts the weight by employing the collected data set. Then, the AP performs the knowledge-building task. The gradient information obtained from the neural network is adopted by the Tuning pass to adjust the frame size. Finally, the optimal frame size and corresponding throughput are recorded to analyze the results.

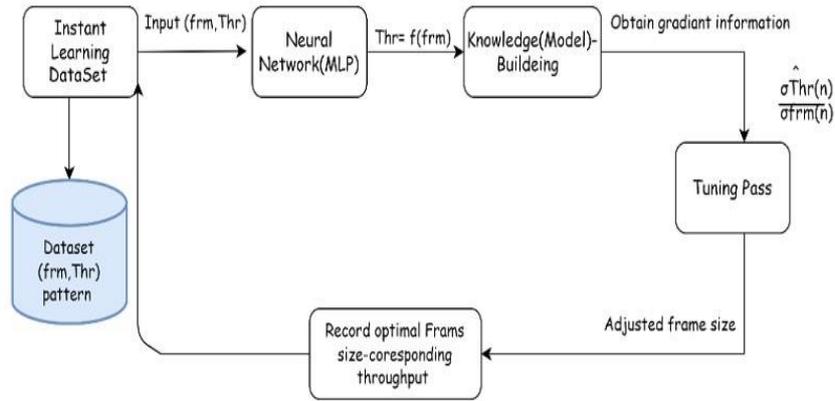

Figure 2. Basic components and flow of the proposed ML model

Table 1. Simulation Parameter and Notation Summary

| Parameters | Symbol | Value |
|---|---|---|
| # Of Antenna at AP | $N_{Ant}$ | 4 |
| # Of Stations | $Num_{STA}$ | 2–4 |
| VoIP traffic payload size |  | 100Byte |
| Video traffic payload size |  | 1000Byte |
| Learning Rate | $\eta$ | 0.5 |
| Mean Square Error Threshold | MES | 0.00001 |
| Epoch Threshold |  | 1000 times |
| Activation Function | Sigmoid ($\sigma$) |  |
| Number of training patterns | $n$ |  |
| Indices of neurons in different layers | $i, j$ |  |
| Frame size(input) of n$^{th}$ training pattern | $frm(n)$ |  |
| Target response for neuron j | $Thr(n)$ |  |
| Actual response of the n$^{th}$ training pattern | $\widetilde{Thr}(n)$ |  |
| Synaptic weight in layer l connecting the output neuron of i to the input neuron j at iteration n | $w_{ji}^{(l)}$ |  |
| Weight sum of all synaptic inputs plus bias of neuron j in layer l at iteration n. | $v_j^l(n)$ |  |
| Signal of output of neuron j in layer l at iteration n | $y_j^l(n)$ |  |
| Local gradient of neuron j in layer l in the tuning pass of hidden layer | $\lambda_j^l(n$ |  |
| Local gradient of neuron j in layer l in the tuning pass of the output layer | $\lambda_1^2(n$ |  |
| Adjustment rate | $\mu$ |  |

# 4. RESULTS AND DISCUSSION

In this section, we evaluate the performance of the proposed machine learning-based adaptive approach to optimize the system frame size in the WLAN downlink MU-MIMO channel that aims to maximize the system throughput performance by considering the effects of channel conditions, heterogeneous traffic patterns, and number of stations.

## 4.1 Experimental Procedure

The training data set is collected by adopting the simulation environment proposed by [16] as a pattern of "frame size - system throughput". The training data set is collected once every 50 seconds. Thus, 50 samples will be collected for each training in considering different network scenarios such as channel conditions, traffic patterns, and number of stations to train the neural network. The system throughput in the data set is the maximum system throughput values obtained from the maximum system throughput achieved by the adaptive aggregation algorithm in [16]. Similarly, the frame size which is used as the input data set in this experiment represents the average offered traffic load generated in the network to obtain the corresponding output i.e., the target system throughput. The weight is updated using these data following the procedure in the backward pass. Forward and backward passes are iteratively performed until the stopping criteria of Mean Square Error (MES) fall below 0.00001 or when the training epoch exceeds 1000 times. The error threshold and the maximum number of iterations determine the accuracy of the function and the computing cost. Then, the tuning pass is executed to adjust the frame size *frm* by adopting the gradient information from the neural network.

The performance of the proposed approach is evaluated by comparing with the system throughput performance achieved by FIFO (Baseline Approach) which was used as a baseline approach to evaluate the performance of the adaptive aggregation approach in [16]. FIFO (Baseline Approach) is an aggregation algorithm which does not consider adaptive aggregation approach [15,16]. Likewise, in this work, we compare the performance of the proposed machine learning approach denoted as Proposed ML Approach in this experiment with the baseline FIFO (Baseline Approach) obtained from [16]. Moreover, we considered the Maximum Throughput achieved by the adaptive aggregation algorithm in [16] to compare it with the Proposed ML Approach to examine how much the Proposed ML Approach effectively optimized the frame size to the maximum system throughput comparably.

In general, the proposed machine-learning based adaptive approach will be evaluated under the following performance factors. The performance of the proposed ML approach is evaluated under the effects of different traffic models such as Pareto, Weibull, and fBM in Section (4.2). Then the performance of the proposed approach under the effect of channel conditions considering SNR= 3, 10, and 20 dB is evaluated in Section (4.3). The performance of the proposed approach under a varying number of STAs (2,3,4) is evaluated in Section (4.4). Finally, the performance of the proposed ML approach is evaluated in terms of system throughput versus optimal system frame size in Section (4.5). All experiments are conducted with a traffic mix of 50% VoIP and 50% video with a constant frame size of 100 Byte and 1000 Byte, respectively.

## 4.2. Performance Under the Effect of Various Traffic Models

In this experiment, the proposed approach is evaluated under the effects of different traffic models such as Pareto, Weibull, and fBM [16], SNR = 10 dB, and Num$_{STA}$= 4. This experiment demonstrates how heterogeneous traffic patterns affect the optimal throughput performance in the WLAN downlink channel. Table 2 illustrates quantitative comparative results of the average maximum system throughput obtained by optimizing the frame size for the proposed ML approach, the baseline FIFO approaches, and Maximum Throughput under the conditions of different traffic models.

Table 2. Quantitative results achieved by the Proposed ML Approach, Maximum Throughput, and FIFO (Baseline Approaches) for average system throughput performance in Mbps under the effects of different traffic models.

| Comparative Approaches | Traffic Models | | |
|---|---|---|---|
| | Pareto | Weibull | fBM |
| FIFO (Baseline Approach) | 511.3145 | 760.629 | 497.7865 |
| Maximum Throughput | 708.9975 | 820.52775 | 728.33775 |
| Proposed ML Approach | 708.724 | 820.4445 | 728.74575 |

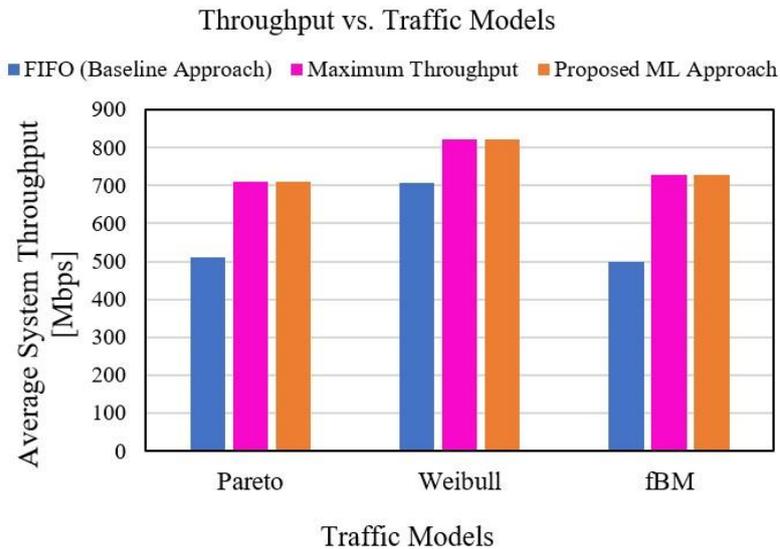

Figure 3. Performance of average system throughput under the effects of heterogenous traffic models when SNR = 10dB.

As the result shows in Figure 3, the proposed ML approach achieved the maximum performance for all traffic models. For instance, for the Weibull traffic model, the maximum performance of 820Mbps is achieved compared to the Pareto and fBM models. Whereas the lowest performance of 708Mbps is achieved for the Pareto traffic model. This indicates that the performance of the proposed ML Approach better copes with the Weibull traffic which is less bursty compared to the other traffic models according to [15]. Thus, these results indicate that traffic patterns in the network determine the system performance. Moreover, the result also demonstrated that the proposed ML approach achieved a compatible result with the maximum system throughput achieved by the adaptive aggregation algorithm i.e., Maximum Throughput proposed by [16]. The FIFO (Baseline Approach) is the worst performance of all traffic models compared to the proposed approach due to its non-adaptive aggregation policy employed in it [15].

## 4.3. Performance Under the Effects of Channel Conditions

In this section the performance of the proposed approach under different channel conditions when SNR = 3, 10, and 20dB, and NumSTA =4 is evaluated as shown in Figure 4 (a), (b), and (c) for the case of different traffic models such as Pareto, Weibull, and fBM. According to the results, the system throughput performance increases when the channel quality improved from 3dB to 20dB better than that of the FIFO (Baseline Approach) due to the adaptive aggregation approach adopted in the Proposed ML Approach. In this regard, the Proposed ML approach achieved the lowest performance of 125Mbps in fBM traffic model as shown in Figure 4 (c), and the maximum of 143Mbps is achieved in the Weibull traffic, when the traffic condition is worst i.e., SNR=3dB. In the contrary, under the near-ideal channel condition, e.g., in SNR of 20dB in the figure, the system throughput performance is almost optimal in all approaches due to lower frame error rate occurred under the near-ideal channel condition. However, the proposed approach achieved the maximum performance of 892Mbps using the Weibull traffic model and the lower 732Mbps is achieved in the Pareto traffic model.

In general, from these results, we can conclude that the performance of the proposed approach is affected by the conditions of traffic patterns and channel conditions. The FIFO (Baseline approach) aggregation policy is the worst compared to the proposed approach in all scenarios because of the non-adaptive aggregation strategy it employs. Moreover, the results also demonstrated that the Proposed ML Approach always archived the maximum performance close to the maximum system throughput achieved by the adaptive aggregation algorithm proposed by [16]. Table 3 illustrates quantitative performance results of the average system throughput performances achieved by the Proposed ML Approach, FIFO (Baseline Approach), and Maximum Throughput under the effects of different channel conditions and traffic models.

Table 3. Quantitative results achieved by the Proposed ML Approach, Maximum Throughput, and FIFO (Baseline Approach) for average system throughput performance in Mbps under the effects of different traffic models and channel conditions.

| Comparative Approaches | Traffic Models | SNR (dB) | | |
|---|---|---|---|---|
| | | 3(dB) | 10(dB) | 20(dB) |
| FIFO (Baseline Approach | | 78.569725 | 511.3145 | 587.10275 |
| Maximum Throughput | Pareto | 139.19 | 708.9975 | 732.4925 |
| Proposed Approach | | 139.21475 | 708.724 | 732.435 |
| | | | | |
| FIFO (Baseline Approach | | 99.8366 | 706.629 | 806.74175 |
| Maximum Throughput | Weibull | 143.7976 | 820.52775 | 892.26925 |
| Proposed Approach | | 143.5908 | 820.4445 | 892.46475 |
| | | | | |
| FIFO (Baseline Approach | | 86.87635 | 497.7865 | 566.02125 |
| Maximum Throughput | fBM | 124.09275 | 728.33775 | 785.82175 |
| Proposed Approach | | 125.05782 | 728.74575 | 786.1 |

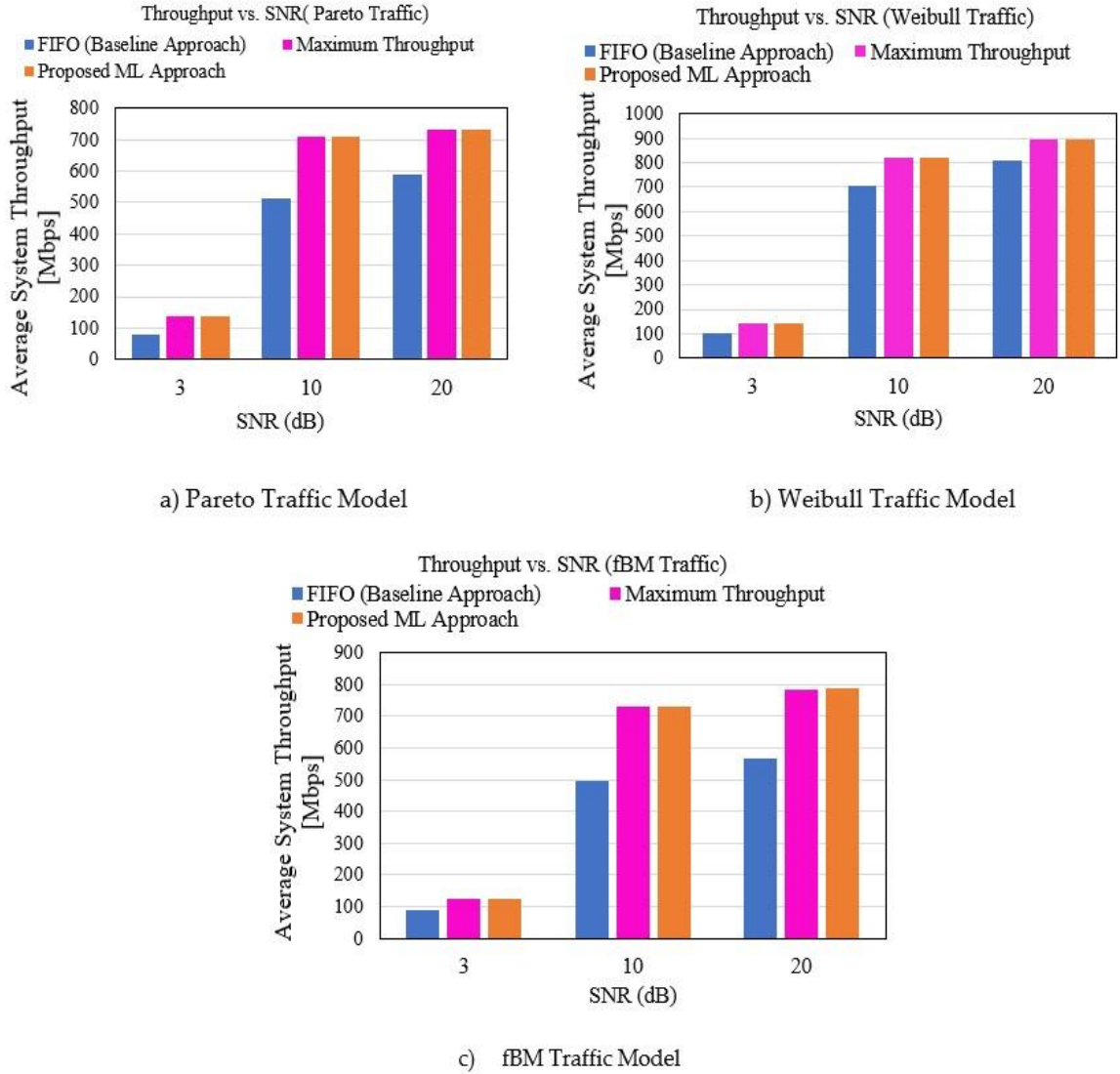

Figure 4. Illustrates System throughput versus SNR for different traffic models such as Pareto, Weibull, and fBM when $Num_{STAs}= 4$.

### 4.4. Performance Under the Effects of Number of Stations

The performance of the proposed approach is evaluated under the effect of different number of stations ($Num_{STA}$ =2, 3, and 4), and when the channel condition is SNR=10dB for the case of Weibull, Pareto, and fBM traffic models. As the results show in Figure 5 (a), (b), and (c), when the number of stations ranges from 2 to 4, the system throughput performance significantly increases in all traffic models as the traffic rate increase with increasing number of stations. However, due to the effect of heterogeneous traffic patterns in different traffic models the performance of the Proposed ML Approach achieved varies even under the same number of stations. Table 4 illustrates quantitative comparative results of the average optimal system throughput achieved by the Proposed ML Approach, Maximum Throughput, and FIFO (Baseline Approaches) under the effects of variable number of STAs.

Table 4. Quantitative results achieved by the Proposed ML approach, Maximum Throughput, and FIFO (Baseline Approach) for average system throughput in Mbps under the effects of variable number of stations in Weibull, Pareto, and fBM traffic models.

| Comparative Approaches | Traffic Models | Number of STAs | | |
|---|---|---|---|---|
| | | 2 | 3 | 4 |
| FIFO (Baseline Approach) | Weibull | 431.467 | 507.87425 | 706.629 |
| Maximum Throughput | | 438.04875 | 620.19375 | 820.52775 |
| Proposed Approach | | 437.902 | 620.18075 | 820.4445 |
| FIFO (Baseline Approach) | Pareto | 338.65 | 357.08 | 511.3145 |
| Maximum Throughput | | 425.56325 | 554.08825 | 708.9975 |
| Proposed Approach | | 425.377 | 553.0695 | 708.724 |
| FIFO (Baseline Approach) | fBM | 311.0525 | 417.13275 | 497.7865 |
| Maximum Throughput | | 396.22 | 566.40875 | 728.33775 |
| Proposed Approach | | 396.0355 | 566.284 | 728.74575 |

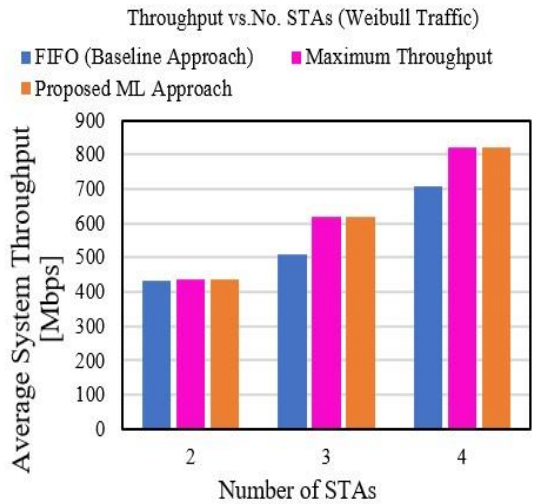
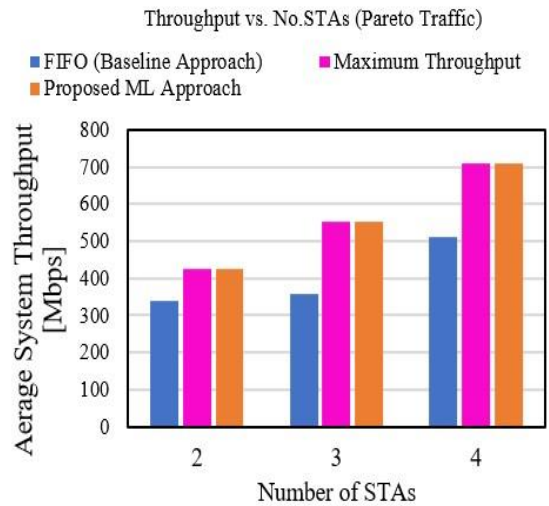

a) Weibull Traffic Model       b) Pareto Traffic Model

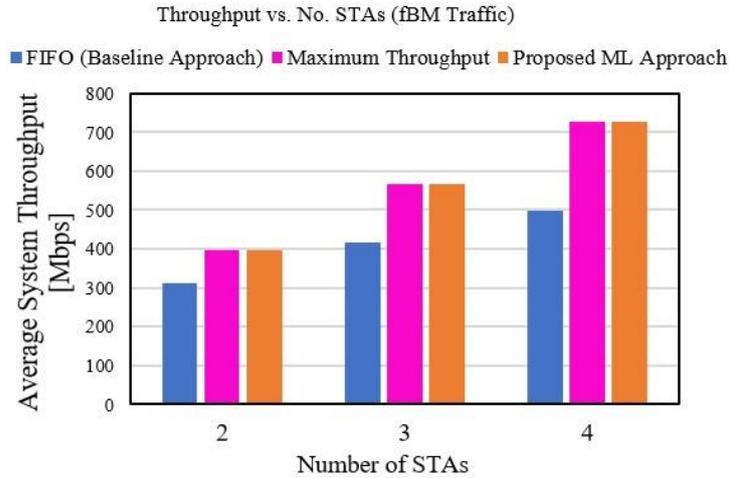

c) fBM Traffic Model

Figure 5. Performance of system throughput versus number of stations when the channel condition is SNR =10dB for the Weibull, Pareto, and fBM traffic models.

As shown in the results in Figure 5, the proposed approach always outperforms the FIFO (Baseline Approaches) in all scenarios due to the adaptive aggregation strategy it adopts. In this regard, the proposed approach achieved the maximum performance of 820Mbps in the case of Weibull traffic whereas the lower performance of 708Mbps is achieved in the Pareto traffic with the same number of STAs. Likewise, when the number of stations equals 2, the worst performance of 396Mbps is achieved by the fBM traffic. These results show that number of stations affects the performance of the system throughput behavior under the conditions of heterogeneous traffic patterns among streams in the downlink MU-MIMO channel. However, the proposed approach always achieved the maximum system throughput performance better than the FIFO (Baseline Approach) closest to the Maximum Throughput of the adaptive aggregation algorithm [16].

**4.5. Performance of System Throughput Vs. Optimal Frame Size**

The results in Figure 6 (a), (b), and (c) show the performance of system throughput behavior with increasing optimal frame size considering SNR= 10 dB, $Num_{STA} = 4$, under the effects of different traffic models Weibull, Pareto, and fBM. This experiment examines the optimal frame size and the corresponding system throughput achieved by the Proposed ML Approach under the effect of different traffic models compared with the FIFO (Baseline Approach).

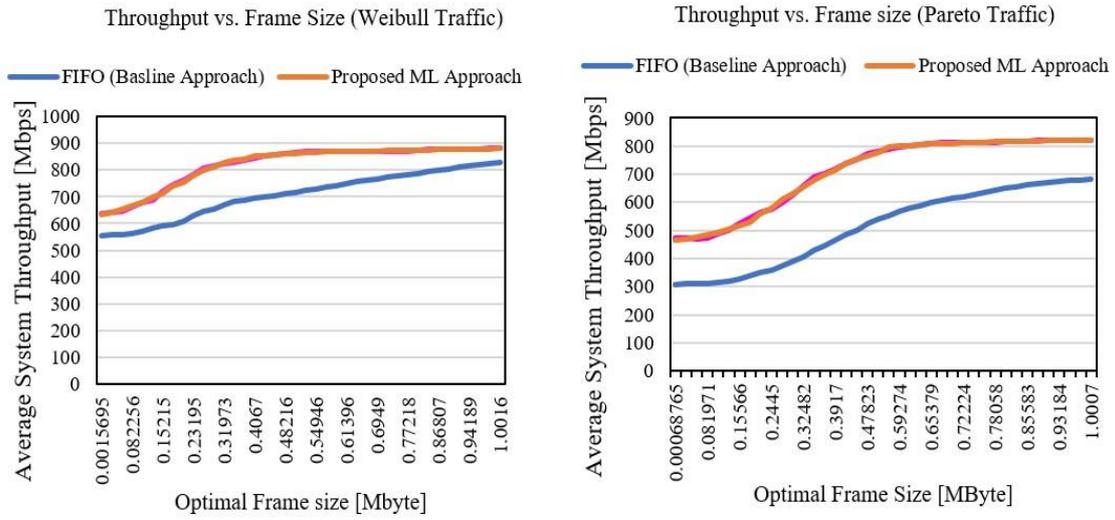

a) Weibull Traffic Model  b) Pareto Traffic Model

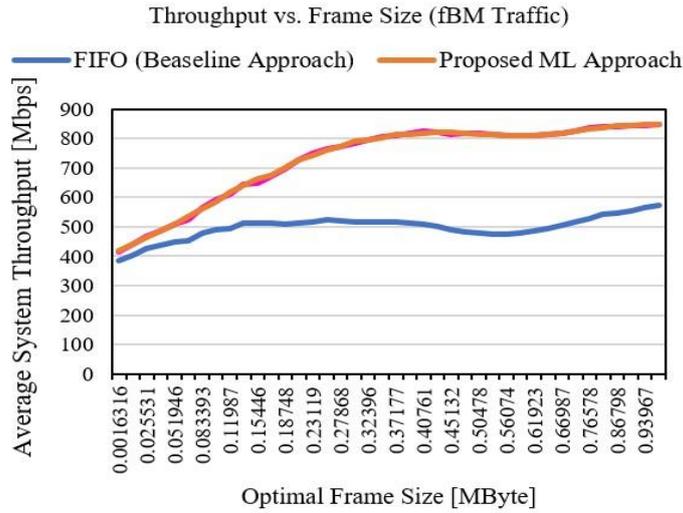

c) fBM Traffic Model

Figure 6. Performance of system throughput versus optimal System frame size when $Num_{STAs} = 4$ and SNR =10dB for the Weibull, Pareto, and fBM traffic models.

According to the results shown in Figure 6 (a), (b), and (c), the proposed ML approach achieved the maximum performance in all traffic models because of the adaptive aggregation approach employed in it in considering channel conditions, traffic patterns, and number of stations. For instance, in the Weibull traffic model, the performance increases with increasing frame size thus achieved the maximum system throughout 880Mbps at the optimal frame size of 1Mbyte. In the case of the Pareto traffic model, the proposed ML approach achieved a maximum system throughput performance of 820Mbps at the optimal system frame size of 1Mbyte. Moreover, as the result shows, the system throughput performance in the Weibull traffic model achieved 630Mbps better than that of the Pareto 470Mbps at the beginning of the result and when the frame size increases. In fBM traffic model the proposed approach achieved the maximum performance of 846Mbps at the optimal system frame size of 0.93Mbyte. However, FIFO Baseline approach achieves the lowest performance in all scenarios because it does not allow adaptive aggregation approach. These results demonstrated that the optimal system frame size achieved is affected by the traffic condition in the network. In this regard, the proposed adaptive ML approach achieved a significant performance by efficiently optimizing the frame size that would maximize the system throughput of WLAN in the downlink MU-MIMO channel taking into account the traffic conditions better than that of the FIFO (Baseline Approach) non-adaptive aggregation approach.

## 5. CONCLUSIONS

The IEEE 802.11n/ac introduced frame aggregation technology to accommodate the growing traffic demand and increases the performance of transmission efficiency and channel utilization. This is achieved by allowing many packets to be aggregated per transmission which realized a significant performance enhancement for the throughput of WLAN. The performance of WLAN depends on different performance factors such as frequency channel, modulation, and coding schemes, transmitter power, etc. at the PHY layer, and retry limit, frame size, contention window size, maximum number of backoffs, etc. at the MAC layer have a significant impact on the performance of WLAN. Optimizing these parameters would improve the system performance of WLAN. Frame size optimization is the main concern of this study. However, it is difficult to efficiently utilize the benefits of frame aggregation in the downlink MU-MIMO channel as stations have heterogeneous traffic demand and data transmission rates. As a consequence, wasted space channel time will occur which degrades transmission efficiency. Moreover, the release of the new IEEE 802.11 standards such as IEEE 802.11ax and IEEE 802.11ay, 5G technologies, and the massive amount of traffic generated in the communication network, allow to expand the set of available communication technologies to compete for the limited radio spectrum resources in pushing for the need of enhancing their coexistence and more effective use of the scarce spectrum resources and to speed up the decision-making process. In response to these performance demands, Machine Learning (ML) is the recent innovating solution to maintain a self-driven network that can configure and optimize itself by reducing human interventions and it is capable of overcoming the drawbacks of traditional mathematical formulations and complex data analysis algorithms. However, most of the existing approaches did not consider a machine-learning-based optimization solution. The main contribution of this paper is to propose a machine-learning-based frame size optimization solution to maximize the system throughput in the WLAN downlink MU-MIMO channel by considering the effect of channel conditions, heterogeneous traffic patterns, and number of stations. In this approach, the AP performs the system throughput measurement and collects the "frame size – throughput" patterns as a data set. To cope with the effects of time-varying channel conditions and heterogeneous traffic patterns, we use online training and iteratively operate the three passes (forward pass, backward pass, and tuning pass) to model the instantaneous *(frm, Thr)* relationship and optimize the frame size. The neural network is used to train these training datasets to accurately model the system throughput with respect to the frame size. Frame size is adjusted according to gradient information which is abstracted from the neural network after the knowledge building. We have performed a simulation experiment to validate that the proposed approach can effectively optimize the system frame size under various channel conditions, traffic patterns, and number of STAs to maximize the system throughput performance of WLAN as compared to the baseline FIFO aggregation algorithm. Moreover, the proposed ML approach can achieve the maximum performance close to the Maximum Throughput of our earlier adaptive aggregation algorithm.

Future work will be conducted by considering the real traffic scenarios. Moreover, the cost of delay and the effects in different channel models such as Rayleigh and Rician on both uplink and downlink WLAN channels will be studied.

## ACKNOWLEDGMENTS

We would like to thank all authors for their contributions and the success of this manuscript. Moreover, we would like to thank the entire authors team for their supportive participation, editors, and anonymous reviewers of this manuscript.

## REFERENCES

[1] IEEE Computer Society. Specific requirements Part 11: Wireless LAN Medium Access Control (MAC) and Physical Layer (PHY) Specifications Amendment 4: Enhancements for Very High Throughput for Operation in Bands below 6 GHz IEEE Computer Society. 2013.
[2] Liao, Ruizhi, Boris Bellalta, Miquel Oliver, and Zhisheng Niu, (2014) "MU-MIMO MAC protocols for wireless local area networks: A survey." *IEEE Communications Surveys & Tutorials,* Vol. 18, No. 1: 162-183.
[3] Yin, Jun, Xiaodong Wang, and Dharma P. Agrawal, (2004) "Optimal packet size in error-prone channel for IEEE 802.11 distributed coordination function." In 2004 IEEE Wireless Communications and Networking Conference (IEEE Cat. No. 04TH8733), Vol. 3, pp. 1654-1659. https://doi.org/10.1109/wcnc.2004.1311801.
[4] Coronado, Estefanía, Abin Thomas, and Roberto Riggio, (2020) "Adaptive ml-based frame length optimization in enterprise SD-WLANs." Journal of Network and Systems Management, Vol. 28, No. 4, pp 850-881.
[5] Kulin, Merima, Tarik Kazaz, Eli De Poorter, and Ingrid Moerman, (2021) "A survey on machine learning-based performance improvement of wireless networks: PHY, MAC and network layer." Electronics Vol. 10, No. 3, pp 318.
[6] Sun, Yaohua, Mugen Peng, Yangcheng Zhou, Yuzhe Huang, and Shiwen Mao, (2019) "Application of machine learning in wireless networks: Key techniques and open issues." IEEE Communications Surveys & Tutorials, Vol. 21, No. 4, pp 3072-3108.
[7] Shea, T. "O and Hoydis J., (2017) " An introduction to deep learning for the physical layer," IEEE Transactions on Cognitive Communications and Networking, Vol 3, No. 4, pp 563-575.
[8] Joo, Er Meng, and Yi Zhou, eds. (2009). Theory and novel applications of machine learning. BoD–Books on Demand.
[9] Luong, Nguyen Cong, Dinh Thai Hoang, Ping Wang, Dusit Niyato, Dong In Kim, and Zhu Han, (2016) "Data collection and wireless communication in Internet of Things (IoT) using economic analysis and pricing models: A survey." IEEE Communications Surveys & Tutorials, Vol. 18, No. 4, pp 2546-2590.
[10] Wang, Cheng-Xiang, Marco Di Renzo, Slawomir Stanczak, Sen Wang, and Erik G. Larsson, (2020) "Artificial intelligence enabled wireless networking for 5G and beyond: Recent advances and future challenges." IEEE Wireless Communications, Vol. 27, No. 1, pp 16-23.
[11] Ci, Song, and Hamid Sharif, (2002) "Adaptive optimal frame length predictor for IEEE 802.11 wireless LAN." In Proceedings of the 6th International Symposium on Digital Signal Processing for Communication Systems (IEE DSPCS'2002).
[12] Bianchi, Giuseppe, (2000) "Performance analysis of the IEEE 802.11 distributed coordination function." IEEE Journal on selected areas in communications, Vol. 18, No. 3, pp 535-547.
[13] Lin, Pochiang, and Tsungnan Lin, (2009) "Machine-learning-based adaptive approach for frame-size optimization in wireless LAN environments." IEEE transactions on vehicular technology, Vol. 58, No. 9 pp 5060-5073.
[14] Modiano, Eytan, (1999) "An adaptive algorithm for optimizing the packet size used in wireless ARQ protocols." Wireless Networks, Vol. 5, No. 4, pp 279-286.
[15] Kassa, Lemlem, Mark Davis, Jingye Cai, and Jianhua Deng, (2021) "A New Adaptive Frame Aggregation Method for Downlink WLAN MU-MIMO Channels." J. Commun. Vol. 16, No. 8, pp 311-322.

[16] Kassa, Lemlem, Mark Davis, Jianhua Deng, and Jingye Cai, (2022) "Performance of an Adaptive Aggregation Mechanism in a Noisy WLAN Downlink MU-MIMO Channel." Electronics, Vol. 11, No. 5, pp 754.
[17] Kim, Sanghyun, and Ji-Hoon Yun, (2019) "Efficient frame construction for multi-user transmission in IEEE 802.11 WLANs." IEEE Transactions on Vehicular Technology, Vol. 68, No. 6, pp 5859-5870.
[18] Bellalta, Boris, Jaume Barcelo, Dirk Staehle, Alexey Vinel, and Miquel Oliver, (2012) "On the performance of packet aggregation in IEEE 802.11 ac MU-MIMO WLANs." IEEE Communications Letters, Vol. 16, No. 10, pp 1588-1591.
[19] Moriyama, Tomokazu, Ryo Yamamoto, Satoshi Ohzahata, and Toshihiko Kato, (2017) "Frame aggregation size determination for IEEE 802.11 ac WLAN considering channel utilization and transfer delay." ICETE 2017 - Proc 14th Int Jt Conf E-Bus Telecommun. Vol. 6, pp 89–94.
[20] Lin, Chi-Han, Yi-Ting Chen, Kate Ching-Ju Lin, and Wen-Tsuen Chen, (2018) "Fdof: Enhancing channel utilization for 802.11 ac." IEEE/ACM Transactions on Networking, Vol. 26, No. 1, pp 465-477.
[21] Lin, Chi-Han, Yi-Ting Chen, Kate Ching-Ju Lin, and Wen-Tsuen Chen, (2017) "acPad: Enhancing channel utilization for 802.11 ac using packet padding." In IEEE INFOCOM 2017-IEEE Conference on Computer Communications, pp. 1-9.
[22] Nomura, Yoshihide, Kazuo Mori, and Hideo Kobayashi, (2016) "High-Efficient Frame Aggregation with Frame Size Adaptation for Downlink MU-MIMO Wireless LANs." IEICE Transactions on Communications, Vol. 99, No. 7, pp 1584-1592.
[23] Haykin, Simon, (1999). Neural Networks: A Comprehensive Foundation. Englewood Cliffs, NJ: Prentice-Hall.
[24] Snyman, Jan A., and Daniel N. Wilke, (2005) Practical mathematical optimization. Springer Science+ Business Media, Incorporated.
[25] Hoi, Steven CH, Doyen Sahoo, Jing Lu, and Peilin Zhao, (2021) "Online learning: A comprehensive survey." Neurocomputing 459 (2021): 249-289.
[26] Online-vs-offline-machine-learning. Available online: https://www.qwak.com/post/ (accessed on 12 June 2022).
[27] Behera, Laxmidhar, Swagat Kumar, and Awhan Patnaik, (2006) "On adaptive learning rate that guarantees convergence in feedforward networks." IEEE transactions on neural networks. Vol. 17, No. 5, pp 1116-1125.

**Authors**

**Lemlem Kassa** has been a senior lecturer in Addis Ababa Science and Technology university in Addis Ababa, Ethiopia since February 2013. She is currently pursuing her PhD study in School of Information and Software Engineering in University of Electronic Science and Technology of China (UESTC) in China. She received the B.S. degree from Micro link Information Technology college in Addis Ababa Ethiopia in 2006 in software engineering and M.S. degree from University Putra Malaysia (UPM), Malaysia in 2009. Her research interests are in the area of wireless communications, artificial intelligence, mobile computing, and software designing.

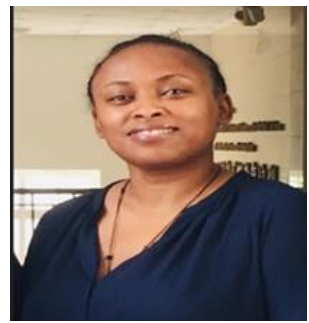

**Dr. Jianhua Deng** graduated in information security from the University of Electronic Science and Technology of China (UESTC), China, in 2006. After graduated, he joined the School of Computer Science and Engineering at UESTC as a staff. From 2009 to 2014, he was a Ph.D. student in Dublin Institute of Technology (DIT) in Ireland and received Ph.D. degree in electrical engineering from DIT in 2014. Now, he is a vice professor in the School of Information and Software Engineering at UESTC. His research interests are in the area of wireless communication, statistical machine learning, artificial intelligence, deep learning. He is the reviewer of some SCI journals (e.g. wireless personal communications).

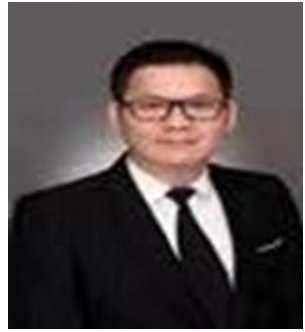

**Prof. Mark Davis** received his BE, MEngSc, and PhD degrees from University College Dublin in 1986, 1989 and 1992 respectively. He is currently the director of the Communications Network Research Institute at Technological University Dublin (TU Dublin). His research interests are in the area of radio resource management techniques for wireless networks, specifically IEEE 802.11 WLANs.

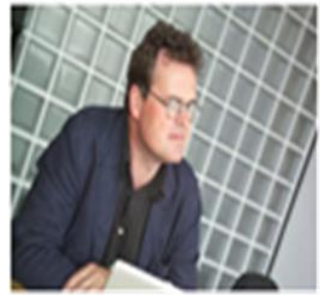

**Prof. Jingye Cai** is Professor and Associate Dean at University of Electronic Science and Technology of China (UESTC). He received BS from Sichuan University in 1983 with a major in radio electronics and Phd from the University of Electronic Science and Technology of China in 1990, with major in signal and information processing. His research interests are in the area of intelligent computing, information engineering, digital Information Processing, and signal processing.

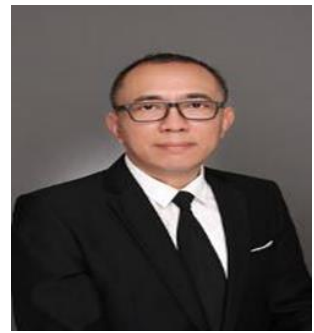